\documentclass[11pt]{article} 
\usepackage{graphicx}
\usepackage{epsfig}
\newcommand{\be}{\begin{equation}}
\newcommand{\ee}{\end{equation}}
\newcommand{\bear}{\begin{eqnarray}}
\newcommand{\eear}{\end{eqnarray}}
\newcommand{\ba}{\begin{array}}
\newcommand{\ea}{\end{array}}

\newcommand{\CL}{{\cal L}} 
\newcommand{\CN}{{\cal N}}

\begin{document}

\begin{titlepage}
\vfill
\begin{flushright}
KIAS-P04050\phantom{abcd}\\
UOSTP-04105\phantom{abcd}\\
{\tt hep-th/0412170}\phantom{ab}\\
\end{flushright}

\vskip 5mm
\begin{center}
{\Large\bf  Separation of Spontaneous Chiral Symmetry Breaking and Confinement via AdS/CFT Correspondence }

\vskip 0.4in

{Dongsu Bak$^a$ 
and  Ho-Ung Yee$^b$}\\
\vskip 12mm

{\it $^a$Physics Department, University of Seoul, Seoul 130-743,
Korea,}\\



\vskip 4mm

{\it $^b$School of Physics, Korea Institute for Advanced Study} \\
{\it 207-43, Cheongryangri-Dong, Dongdaemun-Gu, Seoul 130-722,
Korea}
\\[0.3in]


\end{center}
\begin{center}
({\tt dsbak@mach.uos.ac.kr}, {\tt ho-ung.yee@kias.re.kr} )
\end{center}

\vskip 12mm

\begin{abstract}
\normalsize\noindent

We analyze, in the framework of AdS/CFT correspondence,
the gauge theory phase structure that are supposed to be dual to the recently found
non-supersymmetric dilatonic
deformations to $AdS_5\times S^5$ in type IIB string theory.
Analyzing  the probe D7-brane dynamics in the backgrounds of our interest,
which corresponds to the fundamental $\CN=2$ hypermultiplet,
we show that the chiral bi-fermion condensation responsible for
spontaneous  chiral symmetry breaking is not
logically related to the phenomenon of confinement.


\end{abstract}

\vfill

\end{titlepage}
\setcounter{footnote}{0}

\baselineskip 18pt \pagebreak
\renewcommand{\thepage}{\arabic{page}}
\pagebreak

\section{Introduction }

Quantum Chromodynamics (QCD) is believed to describe hadrons in
the universe. While much of its perturbative dynamics is by now
fairly well understood, it is still hard to analyze, in convincing
ways,  many non-perturbative phenomena that are relevant in low
energy regime. Among these are the confinement of quarks and the
spontaneous breaking of their (approximate) chiral symmetry
(S$\chi$SB). Although these two aspects of QCD have their same
origin in strongly interacting dynamics, there hasn't been found
no logical connection between the two phenomena. In this paper, we
provide, in our belief, one convincing example showing the logical
separation between confinement and S$\chi$SB. Our analysis seems
to suggest that spontaneous chiral symmetry breaking of massless
quarks may happen without any need of a confining potential
between them.

Our analysis is based on the proposal of AdS/CFT correspondence,
in which Type IIB string theory on $AdS_5\times S^5$ background is
equivalent to the $\CN=4$ SYM theory on the boundary of
$AdS_5$\cite{Mal}.
The duality between the two descriptions is supposed to hold even
at the level of Hilbert spaces of their corresponding quantum
theory; (semi-classical) deformations of $AdS_5\times S^5$ which
vanish asymptotically at the boundary correspond to some quantum
states in the dual gauge theory\cite{Wit,Gub1}.
Depending on the deformations in
the bulk that we are considering, these states share several
interesting properties with the usual vacuum states of realistic
gauge theories, such as homogeneity over space and non-vanishing
gluon condensation, etc. Henceforth, studying confinement and
S$\chi$SB on these states may give us an important laboratory for
unraveling the relation between the two phenomena.

The deformed backgrounds of our interest are a family of dilatonic
deformations of $AdS_5\times S^5$ that were found in
Ref.\cite{Bak:2004yf}. A nice fact about these solutions is the
existence of a single adjustable parameter, $k/ \mu$, which
enables us to scan a range of corresponding quantum states. In the
gauge theory side, this parameter represents the ratio of the
gluon condensation to the energy density of the quantum states.
The analysis in Ref.\cite{Bak:2004yf} showed that for ${k/
\mu} <-12$, the potential between (heavy) quark/anti-quark pair is
confining, while states of ${k/\mu}>-12$ were argued to
exhibit Coulomb-like behavior. However, in section 3, we perform a
more careful study for the cases of ${k/ \mu}>-12$ to find the
screening phase instead for them. We also look at the response to
magnetically charged objects and get an interesting phase
structure.

To study S$\chi$SB on these background states, a small number of
light quarks/anti-quarks are introduced a la Karch and Katz in
section 4; probe D7-branes\cite{Kar}.
 They are $N_f$, $\CN=2$
hypermultiplets in fundamental representation of the $SU(N)$ gauge
group, and their effect to $\CN=4$ $SU(N)$ SYM dynamics may be
neglected in $N\gg N_f$ limit via quenched approximation. D7-probe
for studying S$\chi$SB was analyzed first in \cite{Babington:2003vm}, and
its use also for hadron physics \cite{Karch:2002xe,Kruczenski:2003be,Burrington:2004id,Hong:2003jm,Rho:1999jm}
is by now a well
established method in the literature
( See \cite{Sakai:2003wu,Ouyang:2003df,Wang:2003yc,Kruczenski:2003uq,Erdmenger:2004dk} for other set-up's
of introducing flavors.).
From a careful numerical
work, we seem to find a convincing evidence  that S$\chi$SB
persists in the region of our parameter space in which the
confinement no longer exists. Therefore, on the basis of validity
of the AdS/CFT correspondence, it is clear that some states in
large $N$ $\CN=4$ SYM theory, which have non-vanishing gluon
condensation, serve as a rare ground for logical separation
between S$\chi$SB and confinement. We summarize and conclude in
section 5.

\section{Bulk Solutions: Dilatonic Deformation in $AdS_5\times S^5$}

In Ref.\cite{Bak:2004yf}, a family of non-supersymmetric solutions
of type IIB supergravity with asymptotic $AdS_5\times S^5$
geometry were found by turning on generic dilaton deformation to
the maximally supersymmetric $AdS_5\times S^5$
background\footnote{It is also possible to have solutions with the
axion field turned on, but these solutions are readily obtained
from the current ones by $SL(2,Z)$ action. See also
Ref.~\cite{Gut} for the nonsingular class of dilatonic deformation
in $AdS_5$. } (For an earlier example, see \cite{Kehagias:1999tr,Nojiri:1999gf}).
Analytic solutions are available only for the cases
in the Poincare patch, which preserves ${\bf R}\times ISO(3)\times
SO(6)$ subgroup of the full $SO(2,4)\times SO(6)$ symmetry of
$AdS_5\times S^5$. Explicitly, these solutions are

\bear
ds^2&=& 
(y-b)^{1-a\over 4}(y+b)^{1+a\over 4}\left(
-\left(y-b\over y+b\right)^a dt^2 +\frac{dy^2}{16(y-b)^{5-a\over 4}
(y+b)^{5+a\over 4}}+d\vec{x}^2\right)
+d\Omega_5^2
\,,\nonumber\\
\phi&=& \phi_0 +{k\over 8b}\log\left(y-b\over y+b\right)\,,\quad\quad
  F_5= Q
\left(\omega_5+\ast \omega_5\right)\quad, \label{bulkmetric} \eear
where the metric is in the Einstein frame and we let the AdS
radius be unity for simplicity. Here $Q$ is the constant that
counts the number $N$ of D3-branes, $d\Omega_5^2$ and $\omega_5$
are the metric and the volume form of  unit five sphere,
respectively. Clearly, the $S^5$ part of the original $AdS_5\times
S^5$ is intact and the $SO(6)$ R-symmetry of the ${\cal N}=4$ SYM
theory is unbroken at this level. The parameters $a$ and $b$ are
defined in terms of two quantities, $k$ and $\mu$; \be a\equiv
\left(1+{k^2\over 6\mu^2}\right)^{-{1\over 2}}\quad,\quad b\equiv
{\mu\over 2}\left(1+{k^2\over 6\mu^2}\right)^{1\over 2}\quad. \ee
The solutions have a time-like naked singularity at $y=b$. Up to
over-all scaling, these solutions are parameterized by essentially
a single variable $k/ \mu$. They can be thought of as
describing some quantum states in the bulk $AdS_5$ spacetime,
because their deformations to the maximally supersymmetric
$AdS_5\times S^5$ solution decay sufficiently fast as we approach
the boundary. According to AdS/CFT correspondence, we therefore
interpret them as the dual geometries of some quantum states of
the $\CN=4$ SYM gauge theory living on the boundary $R^{1,3}$.

An element of the standard AdS/CFT dictionary gives us an important
information about these quantum states in the gauge theory.
In terms of the coordinate $r$ defined by $r^2=\sqrt{b\over 2}\,e^s$
and $y=b\,\cosh(2s)$, the bulk metric goes to the standard $AdS_5\times S^5$ metric
for large $r$, and $r$ becomes the usual radial coordinate of asymptotic
 $AdS_5$. The dilaton field then asymptotes to
\be
\phi\,\,=\,\,\phi_0+
{k\over 8b}\log\left({y-b\over y+b}\right)\,\,\sim\,\, \phi_0
-\frac{k}{4}\,
\frac{1}{r^4}\quad,
\ee
which implies that the corresponding quantum states in the gauge
theory have a non-vanishing
expectation value of $\CL_{\rm CFT}\sim {1\over 2g_{YM}^2}{\rm tr}\, F^2$ ;
\be
\langle \CL_{\rm CFT} \rangle\,\,=\,\,{k\over 4}\quad.
\ee
The ADM energy density of these states was calculated to be proportional to $\mu$.

\section{Phases of Dual $\CN=4$ SYM States : Confinement vs
Screening}

The family of supergravity backgrounds in the previous section
with varying dilaton profile are supposed to describe some quantum
states of $\CN=4$ SYM theory on $R^{1,3}$. According to AdS/CFT
dictionary, these states are characterized by expectation values
of ${\rm tr}\, F^2$ and the Hamiltonian density. Roughly, we have
seen that
\bear k& \sim & {1\over 2 g^2_{YM}}\left<\,{\rm
tr}(F^2)\,\right>\,\,=\,\, {1\over 2 g^2_{YM}}\ \left<{\rm tr}
({\vec E}^2
-{\vec B}^2)\right>\quad,\nonumber\\
\mu &\sim & {1\over 2 g^2_{YM}}
\left<{\rm tr}
({\vec E}^2
+{\vec B}^2)\right>
\,\,=\,\,{\cal E}\quad, \eear where we denote the energy density
by ${\cal E}$. Though these states are quantum states of the
superconformal $\CN=4$ SYM theory, they have certain properties
that mimic those of interesting vacuum states of more realistic
gauge theories; they are homogeneous over spatial $R^3$ and have
non-vanishing gluon condensation. The latter property has long
been suspected of being one of the crucial characteristics of QCD
vacuum \cite{shifman}. It is thus a meaningful endeavor to study
quantum structure of these states and talk about their ``phases".
One has to bear in mind that the strength of the gluon condensate
here characterizes the macroscopic states of the ${\cal N}=4$ SYM
theory and works as a tunable parameter.

One of the key aspects of a given phase of gauge theory is how
it reacts to external charges.
In the screening phase, external charges are compensated by
conducting currents, and subsequently
screened within some characteristic length scale. Equivalently, the gauge
boson gets massive and does not propagate
beyond its mass scale.
On the other hand, confining phase does not break gauge symmetry
and charge conservation. Instead, electric flux is confined to a narrow string,
resulting in a linear potential
between two charges. One of the most profound observations
in gauge theory is that magnetic screening due to condensation
of magnetically charged object leads to
electric confinement and vice versa. However, there is a caveat here, that is,
if there is also a condensation of electrically charged object at the same time,
electric confinement will be ruined.
In this case, the most plausible expectation is that
both electric and magnetic charges are screened.

In this section, we analyze the response of our states of
 $\CN=4$ SYM theory to various types of external charges,
and find some aspects of interesting phases that we discussed in the above.
In the spirit of AdS/CFT correspondence, external charges are
described by stretched strings in the supergravity background
to the boundary \cite{Rey,Maldacena}.
The Wilson line expectation value is obtained from the effective
world-sheet dynamics of the stretched strings in the
supergravity background. For electric charges, the world-sheet
dynamics is dictated by F1 Nambu-Goto action, while magnetic
or dyonic cases will be described by D1-DBI action with/without
world-sheet gauge flux turned on.

\subsection{Electric Confinement/Screening Transition}
Electric confinement in the above dilatonic backgrounds
has been shown to occur in Ref.~\cite{Bak:2004yf} for  $k/\mu < -12$.
A Nambu-Goto string stretched between heavy quark/anti-quark pair
through the bulk corresponds to the Wilson loop  in the dual
gauge theory side \cite{Rey,Maldacena}. In the large AdS radius limit,
the string behaves classically and one may get the interaction potential
via classical analysis of the Nambu-Goto string dynamics. Here we would like
to analyze more general cases including magnetic
charges as well.

To deal with general $(p,\,\, q)$ string,
let us begin with the Dirac-Born-Infeld action,
\begin{equation}
S= -{1\over 2\pi \alpha'} \int d\tau d\sigma
e^{-\phi}\sqrt{ -\mbox{det}(g_{\mu\nu} \partial_a X^\mu\partial_b X^\nu +2\pi
 \alpha' F_{ab})}\ ,
\end{equation}
where $g_{\mu\nu}$ is the string frame metric which is related to the
Einstein frame metric by
\begin{equation}
g_{\mu\nu}= e^{\phi\over 2} \, g^E_{\mu\nu}\,.
\end{equation}
Denoting
\begin{equation}
M= -\mbox{det}(g_{\mu\nu} \partial_a X^\mu\partial_b X^\nu)\,,
\end{equation}
the Lagrangian density may be written as
\begin{equation}
{\cal L}= -{1\over 2\pi \alpha' e^{\phi}} \sqrt{M -(2\pi \alpha' \, E)^2}  \,,
\end{equation}
where $E= F_{01}$. Let us introduce
the displacement $D$ by
\begin{equation}
D={\partial{\cal L}\over \partial E}={2\pi\alpha' \, E\over e^{\phi}
 \sqrt{M -(2\pi \alpha' \, E)^2}
} \,.
\label{electric}
\end{equation}
$D$ is conserved and ${\partial_\sigma D}=0$ ; one may obtain an
equivalent description of the system by the
Legendre transformation,
\begin{equation}
{\cal L'}=-D\cdot E + {\cal L} =
-{1\over 2\pi \alpha' e^{\phi}} \sqrt{M (1+e^{2\phi}  D^2)}
\end{equation}
by eliminating $E$ using
(\ref{electric}). The displacement  $D$ counts the number of fundamental strings immersed and it is
quantized to take an integer value, which we denote
as $p$. Using the Einstein frame metric,
the above Lagrangian density may  be written as
\begin{equation}
\CL= -{1\over 2\pi \alpha'} \int d\sigma
\sqrt{ -\mbox{det}(g^E_{\mu\nu} \partial_a X^\mu\partial_b X^\nu)}\sqrt{p^2 e^\phi
+q^2 e^{-\phi}
}\quad,
\label{pqlag}
\end{equation}
where we also introduced an integer q counting the number of D-strings. The
derivation
of the $(p,\,\, q)$ string action here is only for $q=1$, but we
generalize it for
an arbitrary $q$. From the above action for the $(p,\,\,q)$ string,
the S-duality of the IIB
string theory is manifest. Namely, the above is invariant under the transformation,
\begin{equation}
g'_{E\mu\nu }=g_{E\mu\nu},\ \  \phi'= -\phi,\ \  p \ \leftrightarrow \  q\,.
\end{equation}
Note that in our dilaton-deformed solutions, the S-duality corresponds to
simply changing $k\  \rightarrow\  -k$ and
$\phi_0\  \rightarrow\  -\phi_0$.
>From this S-duality transformation,
it is clear that magnetic charges are confined for
$k/\mu > 12$, as  electrically charged quarks are confined
if $k/\mu< -12$.

To see the details of the interaction and the phase structure,
let us assume that the $(p,\,\, q)$ string  is static and choose
the gauge $\tau=t$ and $\sigma=y$.
We shall consider the case where the $(p,\,\,q)$ string trajectory is
independent of
$x_2$ and $x_3$. The $(p,\,\,q)$ string Lagrangian then becomes
\begin{equation}
\CL= - \sqrt{\lambda}\int dy \sqrt{ A(y)\left(B(y)+ C(y)\left(
{dx/ dy}
\right)^2\right)}\ ,
\end{equation}
where $\lambda=g_{YM}^2 N$ is the t'Hooft coupling and 
\begin{eqnarray}
A(y) &=&  (y-b)^{\frac{1+3a}{4}}
(y+b)^{\frac{1-3a}{4}}
\left(p^2 e^{\phi_0}
\left({y-b\over y+b} \right)^{\frac{k}{8b}}
+ q^2
e^{-\phi_0}
\left({y-b\over y+b}\right)^{-\frac{k}{8b}}
\right)
\ , \nonumber\\
B(y) &=&  {1\over  16(y-b)(y+b)}\ , \\
C(y) &=& (y-b)^{\frac{1-a}{4}}
(y+b)^{\frac{1+a}{4}}\ . \nonumber
\end{eqnarray}
The computation
showing heavy quark confinement follows closely the one in Ref.~\cite{Bak:2004yf}.
The equation of motion,
\begin{equation}
{d\over dy}\left({ \sqrt{A}\, C\, {dx/dy} \over \sqrt{ B+ C\left(
{dx/ dy}
\right)^2}
} \right)=0\,,
\end{equation}
may be integrated once, and one gets
\begin{equation}
{ \sqrt{A}\, C\, {dx/dy} \over \sqrt{ B+ C\left(
{dx/ dy}
\right)^2}
}=\pm q^{-2}\ ,
\label{velo}
\end{equation}
with an integration constant $q^2$.
To understand the dynamical implication,
we rewrite (\ref{velo}) into the form
\begin{equation}
\left({dy\over dx}\right)^2+ {V}(y)=0\ ,
\end{equation}
with the potential
\begin{equation}
{ V}(y)={C\over B} (1-q^4 AC)
\ .
\end{equation}
This can be viewed as a particle moving in one
dimension
under the potential ${V}$,
regarding the coordinate $x$ as the `time'.

The confinement occurs when the 
`particle' spends
an arbitrarily large `time' when it approaches the
turning point denoted by $y_0$. At the turning point, one has $dy/dx=0$ and
thus ${V}(y_0)=0$, which implies that
\begin{equation}
q_0^4 A(y_0)C(y_0)=1\ ,
\end{equation}
for an appropriate choice of the integration constant $q=q_0$.
The condition of spending arbitrarily large `time'
is fulfilled if ${ V'}(y_0)=0$. This leads to
\begin{equation}
( A C)'|_{y=y_0}=0
\ ,
\end{equation}
where the condition ${ V}(y_0)=0$ is used.

Let us first consider the case of
$(1,\, 0)$ string connecting electrically
charged quark/anti-quark pair.
In this case, the latter condition is solved by
\begin{equation}
y_0 =- ab- k/4
\ .
\end{equation}
For the existence of
the solution in the range $y \in (b,\infty)$, one has to impose
\begin{equation}
y_0-b = -ab- k/4-b \equiv 2b \beta> 0\quad,
\end{equation}
which is equivalent to
\begin{equation}
{k\over \mu} < -12\quad.
\label{wilsoncon}
\end{equation}
Then ${ V}(y_0)=0$ is satisfied by choosing the
integration constant $q$ as
\begin{equation}
q_0^4= {1\over 2b}\beta^\beta (1+\beta)^{-(1+\beta)}
e^{-\phi_0}\ .
\end{equation}

For small $q$, the separation between the quark/anti-quark pair is of
the order of $q$ according to the IR/UV relation.
The energy scale here is much higher than that of the
confinement. Thus the quark/anti-quark potential for sufficiently small separation is of Coulomb type
as expected.



When $\beta>0$ and $q$ approaches $q_0$ from below, the string spends
more and more \lq time' near the turning point $y\sim y_0$.
The separation between the quark and anti-quark becomes larger and larger
as one sends $q$ to $q_0$ from below, because the `time' spent near
the turning point increases more and more. 

In the limit $q \to q_0$, we can compute the tension of the string and
the energy scale of confinement. The energy of the string is given by
\begin{equation}
E_s=   \sqrt{\lambda}\int dy \sqrt{ A\left(B+ C\left(
{dx/ dy}
\right)^2\right)}
= \sqrt{\lambda}\int dx \sqrt{q^4 A^2 C^2}\ ,
\label{qenergy}
\end{equation}
where we have used the equation of motion. The integral in fact diverges
and one may regulate it by subtracting the self-energy of quark and anti-quark.

Since  $q_0^4 A(y_0) C(y_0)=1$ and  the string stays
near the turning point
for most of the \lq time', we find from (\ref{qenergy}) the tension of
the confining string to be
\begin{equation}
T_{QCD}= \sqrt{\lambda}\, \sqrt{A(y_0)C(y_0)}= \sqrt{\lambda}\, q_0^{-2}
= \sqrt{\lambda\,\mu}\,\, {(1+\beta)^{1+\beta\over 2}
\over \sqrt{a}\,\, \beta^{\beta\over 2}}e^{\phi_0\over2}\quad.
\end{equation}
This sets the scale of
confinement. Our result agrees with the previously calculated one in the $\mu\to 0$ limit \cite{Gub}.


\subsection{Screening}

In the analysis of Ref.~\cite{Bak:2004yf}, the region of $k/\mu > -12$
corresponding to $\beta < 0$ was not carefully analyzed because the paper mainly
concerned only about the existence of confinement phenomena.
We would like to show that this region corresponds,  in fact, to the screening phase.
In this region the potential $V$ always has a turning point beyond which the
singularity is located. This feature of inaccessibility to the singularity is true for all value of the integration
constant $q^2$. One may ask the following. May an infinitely large
separated quark/anti-quark pair be connected through this string solution, by 
adjusting
the integration constant $q$? The answer turns out to be no.
Namely, there is an upper limit on the separation length between the
quark and anti-quark pair in the above solutions of string configuration.

To show this, let us first note that
the separation length is given by
\begin{equation}
L=2 \int^\infty_{y_\star} dy {1\over\sqrt{-V}}=
2 \int^\infty_{y_\star} dy
{ \sqrt{B}\over \sqrt{C} \sqrt{q^4 AC-1 }
}\ ,
\end{equation}
where $y_\star$ is the turning point.
For small $q$ satisfying
the condition $e^{\phi_0} \, b\, q^4 \ll 1$, the turning point
$y_\star \sim 1/(e^{\phi_0} q^4)$ is much larger than $b$
and, thus,
the potential $V$ may well be approximated by
\begin{equation}
V \sim 16 y^{5\over 2} (1-e^{\phi_0} q^4 y)\,.
\end{equation}
Then the separation is approximately given by
\begin{equation}
L \sim {e^{\phi_0\over 4} q\over 2} \int^\infty_{1} {dt\over t^{5\over 4}}
{ 1\over\sqrt{t-1 }}
\quad.
\end{equation}
This is the case of sufficiently small separation and  the expression for the separation
is essentially  same as the one for the strings in the pure AdS case, because
the strings are staying in the near boundary
region of the asymptotically AdS space.

To study the upper limit on the separation distance, one should look
at the large $q$ behavior of $L$. When
$e^{\phi_0}\, b\, q^4 \gg 1$, the contribution of the integral
from infinity to $y-b=O(b)$ is of order $1/(e^{\phi_0 \over 2}q^2)$,
which is small. The turning point occurs in the regime
$y-b \ll b$, and the contribution from near the turning point reads as
\begin{equation}
\delta L
= 2 \int_{z_\star} {dz\over 4z^{5-a\over 8} (2b)^{5+a\over 8}}
{1\over \
\sqrt{
e^{\phi_0} q^4 z^{|\beta|} (2b)^{1+\beta}-1
  }
}\,.
\end{equation}
Thus we conclude that
\begin{equation}
\delta L
\sim  q^{-{3+a\over 2|\beta|}}
\quad,
\end{equation}
which is negligible in large $q$ limit.
Obviously, the intermediate region contributes only of order one to the integral.
This shows that the separation has a maximum value for some $q$, which
we denote by $L_{max}$.

\begin{figure}[htb]
\vskip .5cm
\epsfxsize=3.6in
\centerline{
\epsffile{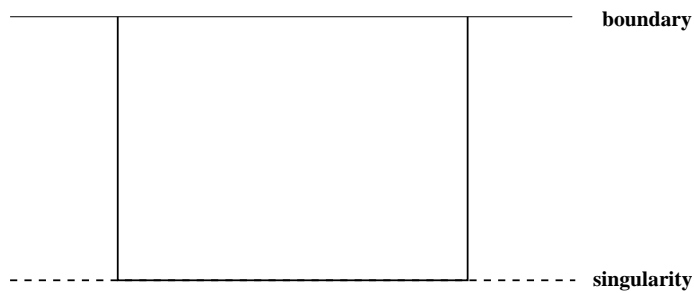}
}
\vspace{.1in}
\caption{\small A string configuration touching the singularity represents
 the screening.}
\label{sc}
\end{figure}

What really happens if the separation of the external quarks becomes larger than
$L_{max}$? In this case, the strings follow the trajectory of the trivial
solution, ${dx\over dy}=0$. The strings from the boundary quarks
and antiquarks are stretched  straight toward the singularity without
any change of $x$
coordinate. Very near the singularity corresponding to IR regime of
the dual field theory, the strings are joined by changing $x$
coordinate. This situation is depicted in Figure \ref{sc}.
One may worry about the part of the string very near the singularity.
(This part describes the physics of the field lines
in the extreme IR regime of the energy scale.). However one may see that
the contribution to the energy of this part is zero at any rate.
To see this, let us first note that the energy integral
of the configuration is given by
\begin{equation}
E_s= \sqrt{\lambda}\int \sqrt{AB dy^2+ AC dx^2}\,.
\end{equation}
Since $dy=0$ for the part of joining two different straight strings,
\begin{equation}
E_{joint}= \sqrt{\lambda}\int \sqrt{AC} dx=  \sqrt{\lambda}\,L\, { (y-b)^{|\beta|\over 2}\phantom{abi}
\over (y+b)^{-{1+\beta\over 2}
}
}\Big|_{y=b}=0\,,
\end{equation}
where in the last equality we have used the fact $\beta < 0$.
Thus the boundary condition at the singularity does not matter 
and the result of vanishing string energy has its own validity despite
the singularity.

From the above discussion, the nature of the interaction  is clear.
The charges are not confined for $k/\mu > -12$. When the separation
becomes larger than $L_{max}$, the quark/anti-quark potential diminishes,
representing the
phenomenon of  screening. Therefore we conclude that
the regime of $k/\mu > -12$ corresponds to a screening phase.

\subsection{Confinement vs Screening of Heavy Quarks}

>From the discussion above, one may expect that the system shows
the phase transition as one varies the parameter
$k/\mu$  by adjusting $k$ or $\mu$. At $k/\mu =-12$, the system undergoes
the phase transition between a confinement phase and a screening phase for heavy
quarks.
The appearance of the tension of the electric-flux string in the confining phase may serve
as an order parameter. At the critical
point of $k/\mu=-12$ or $\beta\ \rightarrow\ 0$
limit, the electric-flux string tension takes a finite value of
\begin{equation}
T_{QCD}=5 \sqrt{\lambda\,\mu}e^{\phi_0\over 2}
\ .
\end{equation}
Namely the tension jumps to the finite value at the phase transition.
This may be understood as follows. Due to the Gauss law, the total electric
flux around charges should remain preserved irrespective of  confinement or
screening. Then, when quarks are confined by the transition,
the electric flux lines form a linear tube
and the finite tension simply comes from the existing energy of the field profile.
Thus the tension should start with a finite value.

\subsection{``Doubly Screening Phase"}
In this subsection, let us consider the response of D-strings
describing the interaction between  magnetically charged
 objects. From the Lagrangian in (\ref{pqlag}), the dynamics of
$(1,\,0)$ strings for a given $k$ is mapped into $(0,\,1)$ strings
with $-k$. Thus, without further computation, one may see that
 magnetically charged quarks are confined when $k/\mu> 12$ and
screened otherwise.

\begin{figure}[htb]
\vskip .5cm
\epsfxsize=3.0in
\centerline{
\epsffile{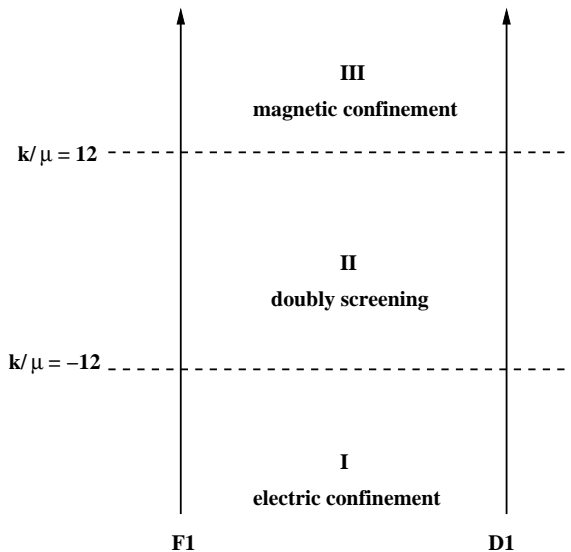}
}
\vspace{.1in}
\caption{\small The full phase diagram.}
\label{phase}
\end{figure}

The full phase structure is drawn in Figure \ref{phase}.
The region I with $k/\mu < -12$ describes the phase where
the electrically charged quarks are confined. Then the
magnetically charged
objects should be screened, 
which
is indeed the case as discussed above.
The region III with $k/\mu > 12$ corresponds to the phase
where  magnetic charges are confined
while quarks are screened. This corresponds to the
S-dual of the region I.

The region II with $-12 < k/\mu < 12$ describes the phase where both
the quarks and magnetic charges are screened, which we call as
`doubly screening phase'. As far as we know, there were no such
examples previously where both the electric and magnetic charges are
screened.
Presumably this phase structure is possible due to the S-duality
symmetry of the underlying $\CN=4$ SYM theory.

\section{Spontaneous Chiral Symmetry Breaking : D7 Probe Analysis}

\subsection{Generalities}

A few $N_f$ D7-branes parallel to a stack of large number
$N$ of D3-branes introduces $N_f$
fundamental $\CN=2$ hypermultiplets in the low energy gauge
dynamics on D3-branes.
For $N$ large enough, $N\gg N_f$, back reaction of the
D7-branes to the near horizon limit of supergravity
background may be irrelevant, and the low energy
gauge dynamics on D3-branes with $\CN=2$ fundamental
hypermultiplets is supposed to be
dual to $AdS_5\times S^5$ with probe D7-branes \cite{Kar}.
The open string dynamics on the probe D7-branes corresponds
to the dynamics of $\CN=2$ hypermultiplets
in the ``ambient" $\CN=4$ SYM theory. This is because the
gauge theory interpretation of the probe approximation
is to take the quenched approximation neglecting effects of
hypermultiplets to the dynamics of $\CN=4$ $SU(N)$ SYM with large $N$.
However, it should be noted that these probe $\CN=2$ fundamental
hypermultiplets experience full dynamics of $\CN=4$ SYM theory.

In the D-brane picture in flat 10-dimensional space-time,
let the D3-brane world-volume span along
$
\{0123\}$,
and the D7-brane along $
\{01234567\}$
directions. The distance between D3 and D7 in the
transverse $\{89
\}$ space gives rise to mass term in the Lagrangian for
hypermultiplets. More specifically, asymptotic value
of $w =\sqrt{x_8^2+x_9^2}$  for large
$\rho^2=x_4^2+x_5^2+x_6^2+x_7^2$ of the D7 world-volume
corresponds to the bare mass $m_f$ of the
hypermultiplets\footnote{We use the notation $x_i=x^i$ for the
spatial coordinates.}.
For the maximally supersymmetric configuration,
$w$ is constant on D7 (say, D7 lies at constant $(x_8,x_9)=(w_0,0)$ while
D3 is sitting at the origin). However, for non-supersymmetric states
such as those we are considering,
$w$ is generically a varying function of $\rho$.

In the supergravity picture, it is not difficult to identify the bare
mass for the hypermultiplets in the framework of
AdS/CFT correspondence.
The maximally supersymmetric supergravity background in the near horizon
limit is $AdS_5\times S^5$ with the string frame metric
\bear
ds^2&=&\frac{1}{f(r)}\left(\sum_{\mu=0}^{3}dx_{\mu}
dx^{\mu}\right)+f(r)\left(\sum_{i=4}^{9}dx^i dx^i\right)\nonumber\\
&=& \frac{r^2}{l^2}\left(\sum_{\mu=0}^{3}dx_{\mu}dx^{\mu}\right)+
\frac{l^2}{r^2} dr^2 +l^2 d\Omega_5^2\quad,\label{adsmetric} \eear
where $r^2=\sum_{i=4}^9 x_i^2=\rho^2+w^2$ and $f(r)=\frac{4\pi N
g_s}{r^2}= \frac{l^2}{r^2}$ is the warping factor. The
world-volume profile of the probe D7-brane in this background is
simply given by identifying the flat D-brane picture coordinate
$\{x^M\}$ ($M=0,\ldots,9$) with the coordinate $\{x^M\}$ in
(\ref{adsmetric}). For example, maximally supersymmetric D7 lying
on the plane $(x_8,x_9)=(w_0,0)$ fills the $AdS_5$ part of
$\{x^{\mu},r\}$ for $w_0\leq r<\infty$, in addition to wrapping
$S^3$-cycle in the $S^5$. The wrapped $S^3$ is defined by
$x_4^2+x_5^2+x_6^2+x_7^2=\rho^2=r^2-w_0^2$ and it vanishes at
$r=w_0$, ensuring a smooth D7 world-volume in $AdS_5\times S^5$.
Note that D7-brane is absent in energy scales below $r=w_0$; this
is consistent with the field theory expectation that we shouldn't
find any hypermultiplet below its mass scale $m_f=w_0$. Moreover,
$w_0$ is a free parameter representing a family of the D7 profiles
; this gives us a freedom of changing the bare mass of
hypermultiplets. Especially interesting limit would be the chiral
symmetry limit\footnote{By chiral symmetry, we mean a chiral
U(1)-symmetry which we discuss more in section (\ref{xsym}).} of
$m_f=w_0=0$.

For the non-supersymmetric backgrounds of our interest, it is possible
to identify suitable coordinate $\{x^M\}$ that
has a natural interpretation of flat coordinate
in the D-brane picture. The string frame metric in this coordinate
is\footnote{From now on, we set $\phi_0=0$ because it plays no special role except
the trivial overall scaling.}
\bear
ds^2= \left(\frac{r^4-1}{r^4+1}\right)^{\frac{k}{8b}}\Bigg\{\!\!\!\!&-&\!\!\!\!
\left(r^2-r^{-2}\right)^{\frac{1+3a}{2}}\left(r^2+r^{-2}
\right)^{\frac{1-3a}{2}}dx^0 dx^0+\frac{1}{r^2}\left(\sum_{i=4}^{9}
dx^i dx^i\right)\nonumber\\
&+&\left(r^2-r^{-2}\right)^{\frac{1-a}{2}}\left(r^2+r^{-2}
\right)^{\frac{1+a}{2}}d \vec{x} \cdot d\vec{x}\Bigg\}\label{newmetric}
\eear
where $r^2=(x_4^2+x_5^2+x_6^2+x_7^2)\,+\,(x_8^2+x_9^2)\equiv\rho^2+w^2$ as before. The above metric is obtained from (\ref{bulkmetric})
by combining $dy^2$ and $d\Omega_5^2$ with a change
of variable $y=b\cosh(2s)$ and $r^2=e^s$. The $R^{1,3}$ coordinate $\{x^0,\vec{x}\}$
has been also rescaled appropriately. The singularity
is now positioned at $r^2=\rho^2+w^2=1$.
The probe D7 world-volume covers $x^0,\ldots,x^7$ and its transverse
position is given by $(x_8,x_9)=(w(\rho),0)$ without loss of generality.
Hence, it fills the (approximate) $AdS_5$ space over $w(0)\leq r <\infty$
(equivalently $0\leq \rho <\infty$), and the wrapped $S^3$-cycle
in $S^5$ shrinks to zero at $r=w(0)$. As we have explained in the previous
paragraph, the field theory
situation dual to this configuration is a quantum state of $\CN=4$ SYM
with $\CN=2$ fundamental hypermultiplet in quenched approximation,
whose bare mass is identified with $m_f=w(\infty)$. A family of profiles
with varying $w(\infty)$ allows us to tune the bare mass,
and in lucky cases, to get the chiral limit.

According to the AdS/CFT proposal, however, the bare mass is not the only information we can extract from $w(\rho)$.
Viewing $w(\rho)$ as an effective scalar field in the $AdS_5$, its asymptotic value at $\rho\rightarrow\infty$
couples to some scalar operator in the field theory. We have actually identified this operator ; we have seen that
$w(\infty)$ couples to the mass operator of the fundamental $\CN=2$ hypermultiplet,
\bear
\delta \CL_{\rm SYM}&=&w(\infty)\int d^2\theta \,\,\tilde Q_f Q_f- ({\rm h.c.})\nonumber\\
&\sim &w(\infty)(\tilde q_L^f q_L^f+{\rm h.c.})+({\rm bosonic})
=w(\infty)\bar q_D^f q_D^f+({\rm bosonic})\,, 
\eear
where we have introduced Dirac fermions,
\be
q_D^f=\left(\ba{c} q_L^f \\ i\sigma^2 (\tilde q^f_L)^* \ea\right)\quad.
\ee
The fermion mass operator is of dimension 3, and the AdS/CFT dictionary tells
us that its expectation value
is encoded in the coefficient of sub-leading $\sim \frac{1}{\rho^2}$ behavior
 of $w(\rho)$ in $\rho\rightarrow\infty$.
Note that the bosonic piece in the above has vanishing expectation value in
the symmetric phase. As we have a freedom of choosing the bare mass $w(\infty)$,
it is possible to see interesting dependence
of bi-fermion mass operator condensate on the bare mass parameter. In
the chiral limit, we may discuss about the occurrence of
spontaneous chiral symmetry breaking.

\subsection{A Subtlety}

In this subsection, we show that the expectation value of the fermion mass operator
for hypermultiplet is precisely given by
the coefficient of sub-leading $\frac{1}{\rho^2}$ in $w(\rho)$
as $\rho\rightarrow\infty$. This is equivalent to a subtle
question of choosing correct field variable, from whose
asymptotic behavior we should read off the expectation value of
the field theory operator. This is a relevant caveat to
care about because $w(\rho)$ has a highly non-standard form of
action functional in $AdS_5$ derived from the D7 DBI action.
The relevant part of the D7-brane DBI action is
\be
S_{\rm D7}=\tau_7\,\int d^8\xi \,e^{-\phi}\sqrt{-
\det\left(\frac{\partial x^M}{\partial \xi^\mu}
\frac{\partial x^N}{\partial \xi^\nu}G^{(10)}_{MN}\right)}\quad,
\ee
where $G^{(10)}$ is the 10-dimensional metric
 of (\ref{newmetric}), and the dilaton profile is
\be
e^{-\phi}=\left(\frac{r^4+1}{r^4-1}\right)^{\frac{k}{4b}}\quad.
\ee
Choosing the gauge $\xi^i=x^i$ ($i=0,\ldots,7$), and
$(x^8,x^9)=(w(\rho),0)$, we obtain the effective action for $w(\rho)$,
\be
S\sim \int d^4x\,\int_0^{\infty}d\rho \,\, \rho^3 \, Z\left(\rho^2+w^2\right)\,\sqrt{1+\left(\frac{d w}{d\rho}\right)^2}\quad,
\label{rhoaction}
\ee
where $Z(x)$ is a complicated function which goes to unity for
large $x$;
\be
Z(x)=\left(1-{1\over x^4}\right)\left({x^2-1}\over{x^2+1}\right)^{k\over{4b}}\quad.
\ee
Another fact that will be important for us later is $Z'(x)\sim
{1\over x^3}$ for large $x$.

For a smooth D7 embedding, we need to impose the boundary condition, ${dw\over d\rho}(0)=0$.
In the asymptotic $\rho\rightarrow \infty$ region, $Z$ goes to unity and the solution of the equation of motion behaves as
\be
w(\rho)\sim m+{C\over \rho^2}\quad.
\ee
Naively, the $\rho$ integration in (\ref{rhoaction}) diverges because for large $\rho$,
\be
\rho^3 \, Z\left(\rho^2+w^2\right)\,\sqrt{1+\left(\frac{d w}{d\rho}\right)^2}\,\,\sim\,\,
\left(\rho^3-\frac{k}{2b}\frac{1}{\rho}\right)+\left(\frac{k}{2b}(2m^2+1)+2C^2\right)\frac{1}{\rho^3}+\cdots
\label{wexpand}
\ee
and we need a suitable regularization procedure.
However, what we are interested in will be variations of the value of (\ref{rhoaction}) under changing $w(\infty)=m$,
and for this purpose, it is enough to regularize (\ref{rhoaction})
by subtracting the value of it at some fixed reference solution $w_0(\rho)$ ;
\be
S_{\rm R}\,\equiv\,\int d^4x\,\int_0^{\infty}d\rho \,\, \rho^3
\left( Z\left(\rho^2+w^2\right)\,\sqrt{1+\left(\frac{d w}{d\rho}\right)^2}
-Z\left(\rho^2+w_0^2\right)\,\sqrt{1+\left(\frac{d w_0}{d\rho}\right)^2}\,\,\right)\quad,
\ee
which is now convergent due to (\ref{wexpand}). The standard AdS/CFT correspondence is then,
\be
\exp\left(i\,S_{\rm R}[m]\right)\,\,=\,\,\left< \exp\left(i\,\int d^4x \,m\,\bar q_D^f q_D^f\right)\right>\quad,
\ee
where $S_{\rm R}[m]$ is the above regularized action evaluated for the solution of the equation of motion with $w(\infty)=m$.
Hence, we have
\be
\frac{\delta S_{\rm R}[m]}{\delta m}= \int d^4 x\,\left<  \bar q_D^f q_D^f\right>\quad.\label{condensate}
\ee

In fact, it is not difficult to calculate the left-hand side of the above relation. Suppose that $w+\delta w$ is the solution
of the equation of motion with $(w+\delta w)(\infty)=m+\delta m$ for infinitesimal $\delta m$. The variation of $S_{\rm R}[m]$ is
\be
\delta S_{\rm R}[m]=\int d^4x\,\int_0^{\infty}d\rho \,\, \rho^3 \,\left(2w\,Z'(\rho^2+w^2)\sqrt{1+\left(\frac{d w}{d\rho}\right)^2}\,\delta w
+Z(\rho^2+w^2){{d w\over d \rho}{d\,\delta w \over d \rho}\over \sqrt{1+\left(\frac{d w}{d\rho}\right)^2}}\right)
\ee
The convergence of this expression may easily be seen from the property $Z'(x)\sim {1\over x^3}$ ,
and we are allowed to perform integration by part
for the second term. The resulting integrand which is proportional to $\delta w$ vanishes because $w(\rho)$ satisfies the equation of motion,
and the surviving surface contribution at $\rho=\infty$ is
\be
\int d^4x\, \lim_{\rho\rightarrow \infty} \left(\rho^3 Z(\rho^2+w^2)
{{d w\over d \rho}\over \sqrt{1+\left(\frac{d w}{d\rho}\right)^2}}\,\delta w\right)=\int d^4x\,(-2C\delta m)\quad,
\ee
using $\rho^3 {d w\over d \rho}\sim -2C$ and $\delta w \sim \delta m$ for large $\rho$.
Comparing with (\ref{condensate}), we thus have
\be
\left<  \bar q_D^f q_D^f\right>\,\,=\,\,-2 C\quad.
\ee

\subsection{The Chiral Symmetry $U(1)_c$ \label{xsym}}

In this subsection, we discuss more about the chiral U(1)-symmetry of the
gauge theory we are considering.
In fact, it is clear in the D-brane picture of D3-D7 system that there must
be a global U(1)-symmetry which
corresponds to the rotation of D7's position in the transverse $(x_8,x_9)$ plane.
(Recall that the D3-branes are aligned along $\{0123\}$, while D7-branes
are along $\{01234567\}$.
Let us put $N$ D3-branes at $(x_8,x_9)=(0,0)$, and $N_f$ D7-branes at
$(x_8,x_9)=(w_1,w_2)$.)
The distance between D3 and D7 introduces a mass term for $\CN=2$
fundamental hypermultiplets,
\be
(w_1+i w_2)\,\int d^2\theta\,\,\tilde Q_f Q^f +{\rm h.c.}\,\,\,\sim\,\,\,
(w_1+i w_2)\, \bar q_D^f q_D^f \,\,+\,\,{\rm (bosonic)}\quad.
\ee
The rotation $(w_1+iw_2)\rightarrow
e^{2i\alpha}\cdot(w_1+i w_2)$ of the D7-brane's position does not change
anything on the D3-brane world-volume in the D-brane picture, and hence
there should exist
a compensating chiral rotation which is a
global symmetry of the gauge theory.
We should also expect the same chiral
symmetry in the non-supersymmetric backgrounds of our interest, because
the solutions preserve the $SO(6)$ symmetry
of $S^5$, which includes $(x_8,x_9)$-plane rotation as a subgroup.

Looking at the superpotential term,
\be
\int d^2\theta\,\,\tilde Q_f Z Q^f\,\,+\,\,\int d^2\theta\,\,
{\rm tr}\left(Z[X,Y]\right)\quad,
\ee
where $X,Y$ and $Z$ are adjoint chiral superfields in $\CN=4$ SYM theory
($Z=X_8+i X_9$),
it is easy to realize that this symmetry is a $R$-symmetry. From the D-brane
picture, we should assign
charges $1$ and $0$ to $Z$ and $X,Y$ respectively. Then $d^2\theta$ has
charge $-1$ (or $\theta_\alpha$ has charge $1\over 2$),
and this forces us to take charge $0$ for $\tilde Q_f$ and $Q_f$.
The reason behind the $R$-symmetry is clear ;
when we rotate D7-branes in the $(x^8,x^9)$ plane, the corresponding
10 dimensional type IIB Killing spinor of D3-D7 system with eight real
components also rotates accordingly.

Note that $\tilde q_f$ and $q_f$  both have charge $-{1\over 2}$
under this $U(1)_c$. In terms of the Dirac spinor $q_D^f$, the
charge is ${1\over 2}\gamma^5={1\over 2} \left(\ba{cc}
-1&0\\0&1\ea\right)$, and a non-vanishing expectation value of
$\left<\bar q_D^f q_D^f\right>$ will break this chiral symmetry
spontaneously. Although $U(1)_c$ has a quantum anomaly which is
proportional to $-N_f\times C_2(F)$, where $C_2(F)$ is the Casimir
invariant of fundamental representation, it is negligible in the
large $N$, t' Hooft limit \cite{Wit1,Wit2}.
From the D-brane
picture, it comes as a surprise that there is an anomaly for
$U(1)_c$ in the effective field theory on D3, because this is a
simple coordinate rotation in the $(x^8,x^9)$-plane. The
resolution of the puzzle lies in the fact that D7-brane sources a
non-trivial profile of RR-scalar $C_0$ around it, such that
rotation in the $(x^8,x^9)$-plane induces a shift monodromy of
$C_0$ field which is exactly proportional to the number of
D7-branes, $N_f$ \cite{Ouyang:2003df}. The RR-scalar $C_0$,
however, couples to the D3-branes by \be C_0\, \int {\rm tr}
(F\wedge F)\quad.\ee Therefore, the shift of the $\theta$
parameter due to the field theory anomaly of $U(1)_c$ rotation is
precisely cancelled by the shift monodromy of the bulk field
$C_0$, and the total anomaly is absent in the whole system. This
may well be called an example of anomaly inflow\footnote{We thank
Jaemo Park for a discussion on this.} (See also \cite{Armoni:2004dc} for a related discussion).

\subsection{Separation Between S$\chi$SB and Confinement}

The equation of motion for $w(\rho)$ from the effective action (\ref{rhoaction}) is somewhat
complicated, and does not seem to have any analytic solutions.
We have performed numerical analysis for solving the equation of motion, and have identified
the asymptotic data, $m$ and $C$, for each solution.
In the previous subsections, we have seen that $m$ corresponds to the bare mass of $\CN=2$ fundamental
hypermultiplets, while $C$ is directly proportional to the condensate of the bi-fermion mass operator for
the hypermultiplets. Hence, a solution whose asymptotic behavior is characterized by $m=0$, but $C\neq 0$,
signals that chiral symmetry is spontaneously broken in the gauge theory living on the boundary.

The effective action (\ref{rhoaction}) for the probe D7-brane  has a parameter
\be
{k\over b}= 2\left(k\over \mu\right)\left(1+{k^2\over 6\mu^2}\right)^{-{1\over 2}}\quad,
\ee
representing a family of bulk type-IIB supergravity backgrounds, which in turn correspond to a family of homogeneous
quantum states of $\CN=4$ SYM theory with $\CN=2$ hypermultiplets in AdS/CFT correspondence.
In section 2, we analyzed ``phases" of these states and observed that their phase structure is sensitive to the value of $k/ \mu$
(equivalently, $k/ b$).
Specifically, for ${k/ \mu }<-12$, ($ {k/ b} <-4.8$), we have an electric confinement, while for
${k/ \mu }>+12$, ($ {k/ b} >+4.8$), magnetically charged objects are confined. An interesting phase seems to
happen for $-12<{k/ \mu }<+12$, ($-4.8 < {k/ b} <+4.8$),
in which both electric charges and magnetic charges are screened.

\begin{figure}[t]
\begin{center}
\scalebox{1.12}[1.2]{\includegraphics{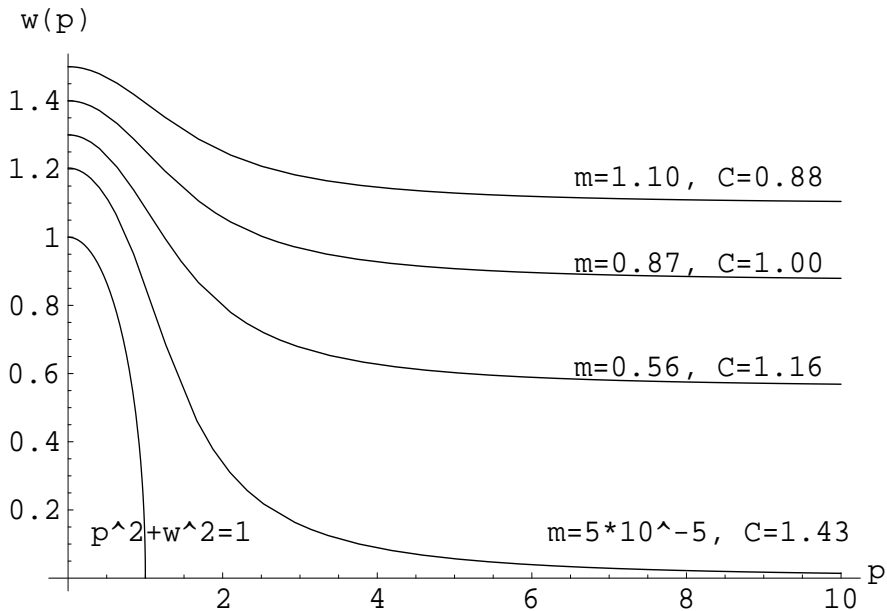}}
\par
\vskip-2.0cm{}
\end{center}
\caption{\small Numerical Solutions for $w(\rho)$ when ${k/ \mu}=-7$ (or
${k/ b}=-4.62$). It is clear that
the solution with $m=0$, but $C\neq 0$, exists. The line $\rho^2+w^2=1$ is the position of the singularity. }
\label{xsb}
\end{figure}

The existence of spontaneous chiral symmetry breaking (S$\chi$SB),
that is, whether there is a solution with $m=0$ but $C\neq 0$ in the bulk,
also depends on the parameter $k/ \mu$ (or $k/ b$). Our numerical study shows that
there is such a solution for ${k/\mu}<-2.97$ (or ${k/ b} <-3.78$). As an exemplar case,
Fig.\ref{xsb} is describing solutions with varying $w(\infty)=m$ when
${k/\mu}=-7$ (or ${k/ b}=-4.62$).
It is evident from the figure that the value of $C$ does not vanish for the solution with $m=0$.
What happens when ${k/ \mu}>-2.97$ (or ${k/ b} >-3.78$) is that solutions start to meet the singularity at $\rho^2+w^2=1$
as we lower the value of $m$. We thus cannot extract useful information for these cases.

The above analysis has a profound implication. For $-12<{k/ \mu}<-2.97$ (or $-4.8<{k/ b} <-3.78$),
the corresponding quantum states of the gauge theory are in the screening phase, while massless fermions of
fundamental representation form a non-vanishing bi-fermion condensation. This contradicts a prevailing lore
that bi-fermion condensation would require a confining potential between two charges.
On the basis of the AdS/CFT correspondence for probe D7-branes, we thus claim to have provided the first example
of separation between spontaneous chiral symmetry breaking and confinement.

\section{Conclusion}

In this work, we have considered dilatonic deformations of AdS geometry that are
dual to some quantum states of the ${\cal N}=4$ SYM theory with non-vanishing gluon
condensation, $k$, as well as homogeneous energy density $\mu$. As varying the parameter
$k/\mu$, we have identified the phases of these states by studying the
interaction between quarks/anti-quarks, and also between magnetically charged objects.
The regime $k/\mu < -12$
is electrically confining, where quarks are confined and magnetic charges are
screened.  The opposite regime of $k/\mu > 12$ corresponds to the S-dual transformed
phase, where magnetic charges are confined. For $-12 < k/\mu < 12$,
interestingly both fundamental quarks as well as magnetic charges are screened, whose phase
 we call ``doubly screening phase''.

We then introduced the probe D7-branes and studied possible spontaneous
chiral symmetry breaking. The ${\cal N}=2$ fundamental hypermultiplet
arising from the D3-D7 strings possesses the classical chiral $U(1)_c$,
which suffers from quantum anomaly. However we are working in the large $N$
limit of D3-branes and the effect of the anomaly may be ignored.
By studying the D7 moduli dual to the fermion mass
operator of the hypermultiplet, we have shown that there is a nonvanishing
 bifermion condensate in the zero-mass limit, leading to the
spontaneous breaking of the chiral symmetry. We demonstrated that this
happens even within the screening phase with no confinement.

It is our hope that the conclusions we have drawn from analyzing these
states of $\CN=4$ SYM theory
reflect some truth of generic confining gauge theories. At least,
it seems to suggest that
spontaneous chiral symmetry breaking does not necessarily require confinement.

\vskip 1cm \centerline{\large \bf Acknowledgement} \vskip 0.5cm


We would like to thank Kimyeong Lee, Jaemo Park and Soo-Jong Rey  
for helpful discussions. We also thank other participants of the
``AdS/CFT and Quantum Chromodynamics" (Oct. 28-30, 2004), Hanyang
University, Korea, for inspiring discussions.
D.B. is supported in part by KOSEF ABRL
R14-2003-012-01002-0 and KOSEF
R01-2003-000-10319-0. D.B. also likes to thank the warm hospitality of KIAS
where part of this work was done.
H.-U.Y. is partly supported
by grant No. R01-2003-000-10391-0 from the Basic Research Program
of the Korea Science \& Engineering Foundation.

 \vfil


\begin{thebibliography}{99} \frenchspacing





\bibitem{Mal}
J.~M.~Maldacena, ``The large N limit of superconformal field
theories and supergravity,'' Adv.\ Theor.\ Math.\ Phys.\  {\bf 2},
231 (1998) [Int.\ J.\ Theor.\ Phys.\  {\bf 38}, 1113 (1999)]
[arXiv:hep-th/9711200].

\bibitem{Wit}
E.~Witten,
``Anti-de Sitter space and holography,'' Adv.\ Theor.\
Math.\ Phys.\  {\bf 2}, 253 (1998) [arXiv:hep-th/9802150].

\bibitem{Gub1}
S.~S.~Gubser, I.~R.~Klebanov and A.~M.~Polyakov,
``Gauge theory
correlators from non-critical string theory,'' Phys.\ Lett.\ B
{\bf 428}, 105 (1998) [arXiv:hep-th/9802109].



\bibitem{Bak:2004yf}
D.~Bak, M.~Gutperle, S.~Hirano and N.~Ohta,
``Dilatonic repulsons and confinement via the AdS/CFT correspondence,''
Phys.\ Rev.\ D {\bf 70}, 086004 (2004)
[arXiv:hep-th/0403249].





\bibitem{Kar}
A.~Karch and E.~Katz,
``Adding flavor to AdS/CFT,''
JHEP {\bf 0206}, 043 (2002)
[arXiv:hep-th/0205236].




\bibitem{Karch:2002xe}
A.~Karch, E.~Katz and N.~Weiner,
``Hadron masses and screening from AdS Wilson loops,''
Phys.\ Rev.\ Lett.\  {\bf 90}, 091601 (2003)
[arXiv:hep-th/0211107].


\bibitem{Kruczenski:2003be}
M.~Kruczenski, D.~Mateos, R.~C.~Myers and D.~J.~Winters,
``Meson spectroscopy in AdS/CFT with flavour,''
JHEP {\bf 0307}, 049 (2003)
[arXiv:hep-th/0304032].\\
N.~J.~Evans and J.~P.~Shock,
``Chiral dynamics from AdS space,''
Phys.\ Rev.\ D {\bf 70}, 046002 (2004)
[arXiv:hep-th/0403279].\\
J.~L.~F.~Barbon, C.~Hoyos, D.~Mateos and R.~C.~Myers,
``The holographic life of the eta',''
JHEP {\bf 0410}, 029 (2004)
[arXiv:hep-th/0404260].


\bibitem{Babington:2003vm}
J.~Babington, J.~Erdmenger, N.~J.~Evans, Z.~Guralnik and I.~Kirsch,
``Chiral symmetry breaking and pions in non-supersymmetric gauge/gravity
duals,''
Phys.\ Rev.\ D {\bf 69}, 066007 (2004)
[arXiv:hep-th/0306018].\\
K.~Ghoroku and M.~Yahiro,
``Chiral symmetry breaking driven by dilaton,''
Phys.\ Lett.\ B {\bf 604}, 235 (2004)
[arXiv:hep-th/0408040].

\bibitem{Burrington:2004id}
 B.~A.~Burrington, J.~T.~Liu, L.~A.~Pando Zayas and D.~Vaman,
``Holographic duals of flavored N = 1 super Yang-Mills: Beyond the
probe approximation,''
[arXiv:hep-th/0406207].




\bibitem{Hong:2003jm}
S.~Hong, S.~Yoon and M.~J.~Strassler,
``Quarkonium from the fifth dimension,''
JHEP {\bf 0404}, 046 (2004)
[arXiv:hep-th/0312071].\\
S.~Hong, S.~Yoon and M.~J.~Strassler,
``On the couplings of vector mesons in AdS/QCD,''
[arXiv:hep-th/0409118].\\
S.~Hong, S.~Yoon and M.~J.~Strassler,
``Adjoint trapping: A new phenomenon at strong 't Hooft coupling,''
[arXiv:hep-th/0410080].\\
M.~Bando, T.~Kugo, A.~Sugamoto and S.~Terunuma,
``Pentaquark baryons in string theory,''
Prog.\ Theor.\ Phys.\  {\bf 112}, 325 (2004)
[arXiv:hep-ph/0405259].



\bibitem{Rho:1999jm}
M.~Rho, S.~J.~Sin and I.~Zahed,
``Elastic parton parton scattering from AdS/CFT,''
Phys.\ Lett.\ B {\bf 466}, 199 (1999)
[arXiv:hep-th/9907126].\\
S.~J.~Sin and I.~Zahed,
``Holography of radiation and jet quenching,''
[arXiv:hep-th/0407215].











\bibitem{Sakai:2003wu}
T.~Sakai and J.~Sonnenschein,
``Probing flavored mesons of confining gauge theories by supergravity,''
JHEP {\bf 0309}, 047 (2003)
[arXiv:hep-th/0305049].\\
S.~Kuperstein,
``Meson spectroscopy from holomorphic probes on the warped deformed
conifold,''
[arXiv:hep-th/0411097].



\bibitem{Ouyang:2003df}
P.~Ouyang,
``Holomorphic D7-branes and flavored N = 1 gauge theories,''
Nucl.\ Phys.\ B {\bf 699}, 207 (2004)
[arXiv:hep-th/0311084].


\bibitem{Wang:2003yc}
X.~J.~Wang and S.~Hu,
``Intersecting branes and adding flavors to the Maldacena-Nunez  background,''
JHEP {\bf 0309}, 017 (2003)
[arXiv:hep-th/0307218].\\
C.~Nunez, A.~Paredes and A.~V.~Ramallo,
``Flavoring the gravity dual of N = 1 Yang-Mills with probes,''
JHEP {\bf 0312}, 024 (2003)
[arXiv:hep-th/0311201].



\bibitem{Kruczenski:2003uq}
M.~Kruczenski, D.~Mateos, R.~C.~Myers and D.~J.~Winters,
``Towards a holographic dual of large-N(c) QCD,''
JHEP {\bf 0405}, 041 (2004)
[arXiv:hep-th/0311270]. \\
T.~Sakai and S.~Sugimoto,
``Low energy hadron physics in holographic QCD,''
[arXiv:hep-th/0412141].

\bibitem{Erdmenger:2004dk}
J.~Erdmenger and I.~Kirsch,
``Mesons in gauge / gravity dual with large number of fundamental fields,''
JHEP {\bf 0412}, 025 (2004)
[arXiv:hep-th/0408113].


\bibitem{Kehagias:1999tr}
A.~Kehagias and K.~Sfetsos,
``On running couplings in gauge theories from type-IIB supergravity,''
Phys.\ Lett.\ B {\bf 454}, 270 (1999)
[arXiv:hep-th/9902125].\\
A.~Kehagias and K.~Sfetsos,
``On asymptotic freedom and confinement from type-IIB supergravity,''
Phys.\ Lett.\ B {\bf 456}, 22 (1999)
[arXiv:hep-th/9903109].


\bibitem{Nojiri:1999gf}
S.~Nojiri and S.~D.~Odintsov,
``Running gauge coupling and quark anti-quark potential from dilatonic
gravity,''
Phys.\ Lett.\ B {\bf 458}, 226 (1999)
[arXiv:hep-th/9904036].\\
S.~Nojiri and S.~D.~Odintsov,
``Curvature dependence of running gauge coupling and confinement in  AdS/CFT
correspondence,''
Phys.\ Rev.\ D {\bf 61}, 044014 (2000)
[arXiv:hep-th/9905200].








\bibitem{Gut}
D.~Bak, M.~Gutperle and S.~Hirano,
``A dilatonic deformation of AdS(5) and its field theory dual,''
JHEP {\bf 0305}, 072 (2003)
[arXiv:hep-th/0304129].



\bibitem{shifman}
M.~A.~Shifman, A.~I.~Vainshtein and V.~I.~Zakharov,
``QCD And Resonance Physics. Sum Rules,''
Nucl.\ Phys.\ B {\bf 147}, 385 (1979).




\bibitem{Rey}
S.~J.~Rey and J.~T.~Yee,
``Macroscopic strings as heavy quarks in large N gauge theory and  anti-de
Sitter supergravity,''
Eur.\ Phys.\ J.\ C {\bf 22}, 379 (2001)
[arXiv:hep-th/9803001].


\bibitem{Maldacena}
J.~M.~Maldacena,
``Wilson loops in large N field theories,''
Phys.\ Rev.\ Lett.\  {\bf 80}, 4859 (1998)
[arXiv:hep-th/9803002].






\bibitem{Gub}
S.~S.~Gubser,
``Dilaton-driven confinement,''
[arXiv:hep-th/9902155].

\bibitem{Wit1}
E.~Witten,
``Current Algebra Theorems For The U(1) 'Goldstone Boson',''
Nucl.\ Phys.\ B {\bf 156}, 269 (1979).


\bibitem{Wit2}
E.~Witten,
``Theta dependence in the large N limit of four-dimensional gauge  theories,''
Phys.\ Rev.\ Lett.\  {\bf 81}, 2862 (1998)
[arXiv:hep-th/9807109].

\bibitem{Armoni:2004dc}
A.~Armoni,
``Witten-Veneziano from Green-Schwarz,''
JHEP {\bf 0406}, 019 (2004)
[arXiv:hep-th/0404248].





\end{thebibliography}
\end{document}